\documentclass[12pt,a4paper,onecolumn]{article}
\usepackage[utf8]{inputenc}
\usepackage{graphicx}
\usepackage{amsmath}
\usepackage{hyperref}
\usepackage{geometry}
\usepackage{float}
\usepackage{subcaption}
\usepackage{newunicodechar}
\newunicodechar{₂}{$_2$}
\title{PhoQuPy: A Python framework for Automation of Quantum Optics experiments}
\author{Srivatsa Murali, Anshuman Kumar \\ Laboratory of Optics of Quantum Materials, IIT Bombay}
\date{January 2026}

\begin{document}

\maketitle
\begin{abstract}
We present the automation of a confocal photoluminescence (PL) scanning system for the 
identification and characterization of single-photon emitters (SPEs) in quantum materials. 
The setup excites the sample with a laser and acquires a spectrum at each spatial 
coordinate in a raster scan pattern. A double-acquisition method is used to remove cosmic ray artifacts by comparing subsequent measurements at the same spatial coordinate. Once 
identified, the emitter is further characterized via a HBT setup, thereby measuring 
lifetime as well as second-order autocorrelation $g^{(2)}$ measurements to confirm single-photon emission. The system integrates Python-based hardware control for motorized stages, spectrometer acquisition, and post-processing, with a migration to a galvo-mirror scanning approach for using it along with a cryostat for low temperature measurements. Our results demonstrate spatially resolved PL maps and temperature-dependent spectra, highlighting the capability of the setup to efficiently benchmark SPE performance. We further went on to perform automation of other experiments such as a Non-Linear Interferometry setup for Quantum Imaging with Undetected Light and a Fourier Transform Imaging Spectroscopy using a common path birefringence Interferometer to obtain hyperspectral images of our samples.
\end{abstract}

\section{Introduction}

The main focus of this report is to automate the optical characterization of quantum materials using techniques like confocal PL mapping, $g^{(2)}$ correlation measurements, stitched imaging, and cryostat-based scanning. We need to combine various instruments, including piezo stages (NanoMax MAX312D), galvano mirrors (ScannerMax Saturn 9B - 10mm), spectrometers (Andor Kynera 328i), CCDs (Newton iDUS), CMOS cameras (Thorlabs CS-165CU), EMCCDs(Princeton Teledyne ProEM HS-512), Nireos Gemini Interferometer and SPADs (Excelitas SPCM-AQRH) .

Most of these characterization techniques are not possible by manual operation hence requiring automation that can offer reliable and repeatable measurements. By connecting motion stages, spectrometers, single-photon detectors, galvo scanners, and DAQ systems within a unified Python framework, we can perform experiments with consistent timing and less human involvement. We then created improved acquisition strategies, such as automated cosmic-ray suppression, adaptive scanning, and synchronized multi-device operation.

Python’s extensive hardware ecosystem, including the APIs for Zaber, Thorlabs, Andor, PicoQuant (snAPI), and NI-DAQ, enables smooth coordination of different instruments. Modular device abstractions handle low-level communication while providing simple high-level commands, which makes the system flexible and easy to maintain and expand.

In the Laboratory of Optics of Quantum Materials (LOQM), we use this framework to study quantum emitters in materials like hBN, SiN, WSe\(_2\), and colloidal quantum dots.\cite{Esmann2024} Setups often need reconfiguration, temperature-dependent measurements, or new optical pathways. A unified automation system greatly simplifies these tasks and improves reproducibility across experiments.

The next sections discuss the structure, implementation, challenges, and future developments of this automation framework.

\section{System Architecture Overview}

The architecture separates experiment logic from hardware-specific implementations. This approach allows for modularity and reuse across different setups.

\begin{figure}[h]
    \centering
    \includegraphics[width=0.75\linewidth]{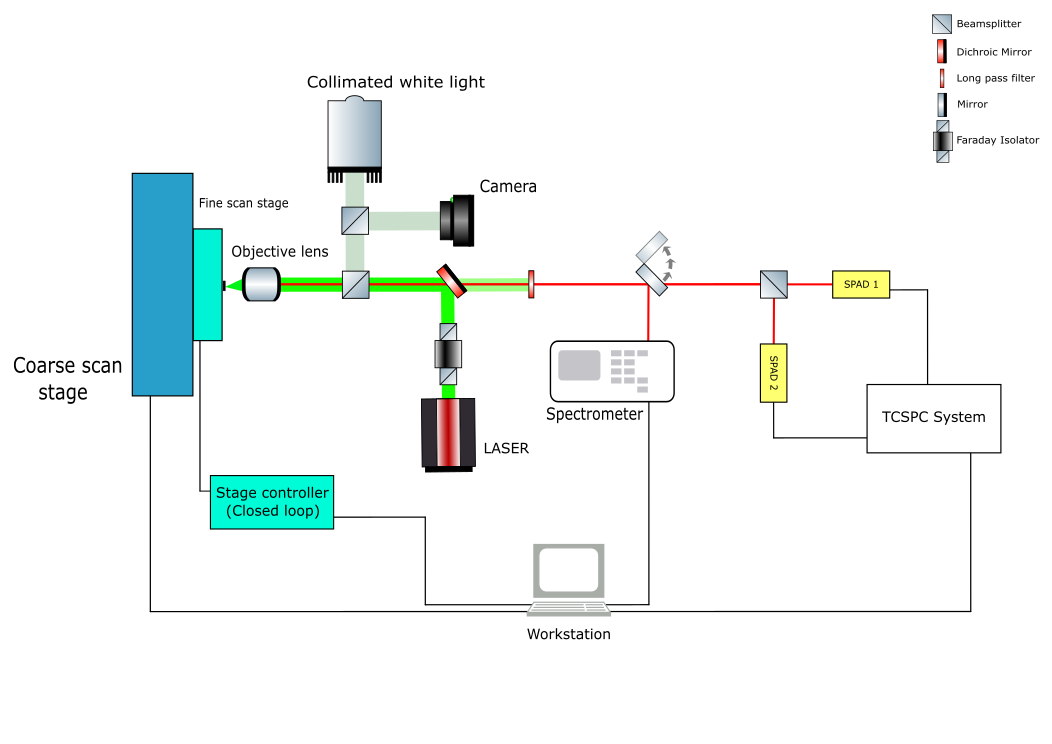}
    \caption{System architecture overview. A modular Python layer abstracts all hardware and coordinates experiment workflows.\cite{KrammPythonAutomation}}
\end{figure}

Laser excitation passes through shaping optics and scanning elements before it reaches the sample. Room-temperature experiments use NanoMax or Zaber stages for movement. In contrast, cryostat experiments rely on ScannerMAX galvo mirrors driven by a Mach-DSP controller. This setup allows for fast, vibration-free optical scanning.

Photoluminescence (PL) emission is collected and directed to an Andor Kymera 328i spectrometer. Automated control through the Andor SDK allows for dynamic adjustments of exposure time, slit width, and grating configurations.

This spectrum is collected at every spatial point in the sample is the region that we need to scan and saved to a 2D array that contains spectrum intensities stored along a column and spectrum at each position is saved in subsequent columns. We then convert this to a map by simply deploying a color-map where the intensity at the selected wavelength at each spectrum is attributed as the intensity at that pixel, and clicking on that pixel would display the full spectrum associated with that pixel beside it as shown in Fig 3.

For single-photon detection, we use SPAD modules connected to PicoQuant TCSPC hardware. snAPI enables timestamp acquisition, channel setup, and synchronization control.

Electronic communication includes serial, USB, and analog input/output.

The Python integration layer handles hardware initialization, scan routines, data transfer, pre-processing, and metadata storage in organized formats. Real-time visualization supports alignment, drift monitoring, and assessment of scan progress.


\section{Joystick Control and Image Stitching Using Zaber Stage}

The Zaber stage allows for easy manual navigation with a joystick and automated image stitching.

Joystick control makes coarse alignment straightforward. The joystick inputs connect to real-time stage speeds, enabling movement within acceleration limits and safety bounds. For stitched imaging, the stage moves along a set grid while taking partially overlapping microscope images. We detect features across the tiles and create a smooth composite image using the MIST algorithm. This process first calculates the translations of each image between its vertical and horizontal neighbors using edge detection. Then, it fine-tunes the image translations based on a mechanical model of the stage movements and assembles the stitched images using a minimum spanning tree.\cite{Chalfoun2017}

\begin{figure}[h]
    \centering
    \includegraphics[width=0.85\linewidth]{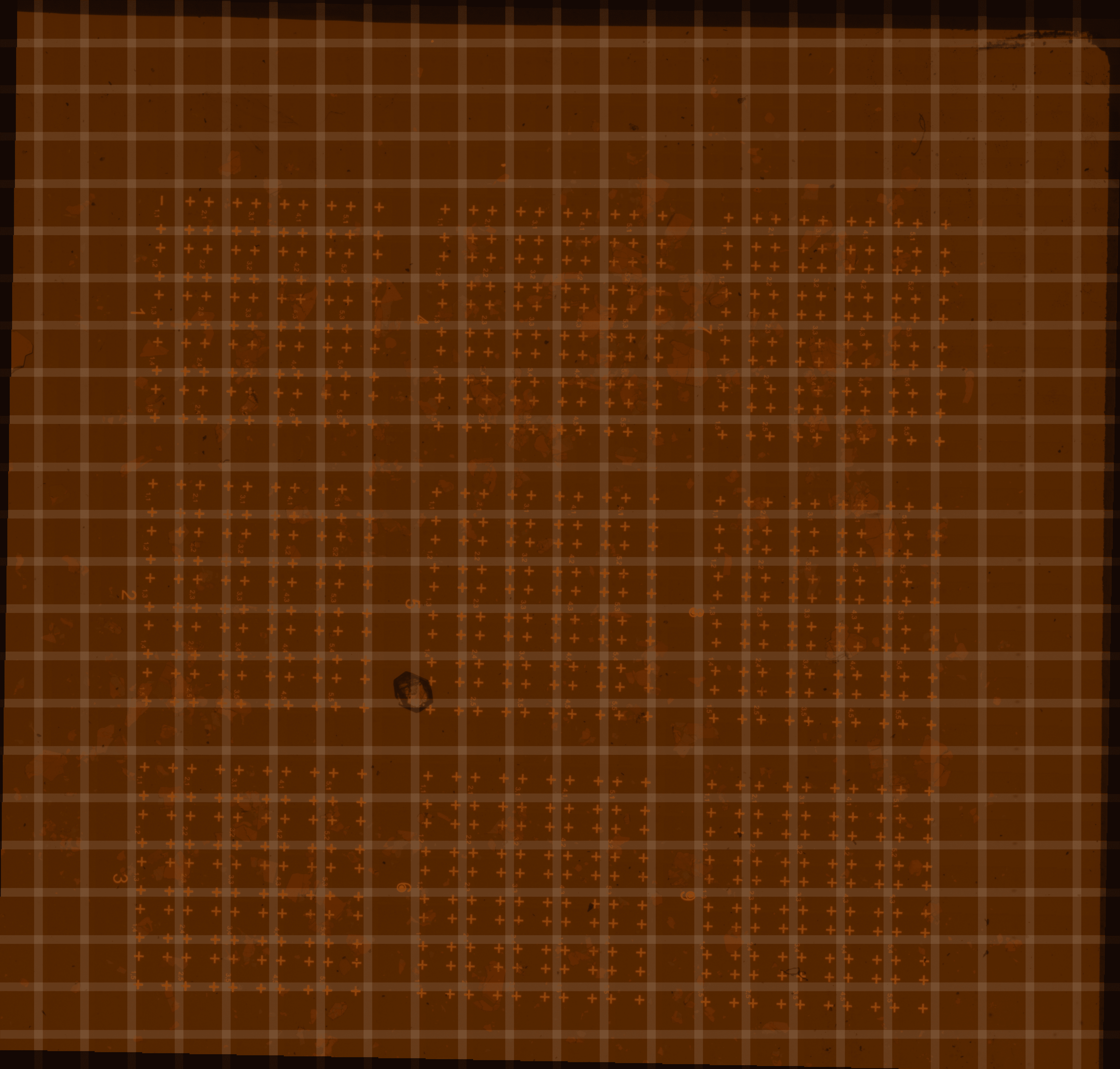}
    \caption{20×20 stitched image covering approximately $4\,\mathrm{cm}^{2}$}
\end{figure}


\section{Confocal Photoluminescence (PL) Mapping Automation}

Confocal PL mapping finds localized emitters by scanning a focused laser beam in a grid pattern while capturing PL spectra. This method improves spatial resolution and reduces background fluorescence.

\begin{figure}[h]
    \centering
    \includegraphics[width=0.8\linewidth]{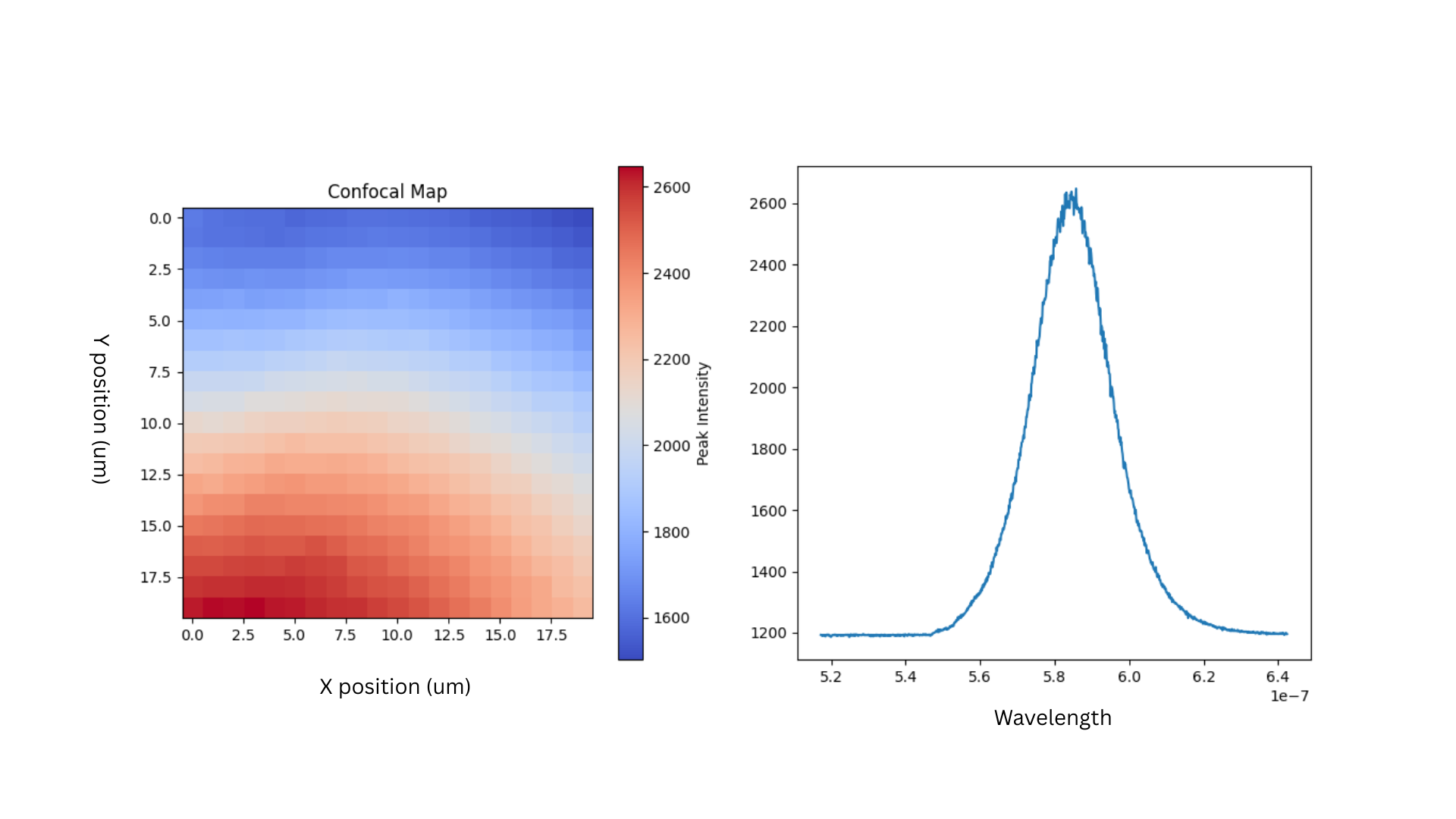}
    \caption{Confocal PL map interface (CdSe quantum dot).}
\end{figure}

\begin{figure}[h]
    \centering
    \includegraphics[width=0.75\linewidth]{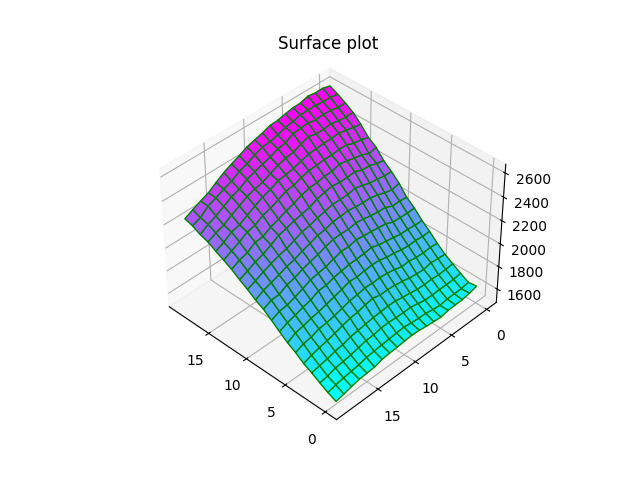}
    \caption{Surface plot of PL intensity map for CdSe Quantum Dot}
\end{figure}

The setup supports mechanical scanning using NanoMax or Zaber translation stages and fast optical scanning through galvo mirrors, especially within a cryostat. The Andor spectrometer records a complete photoluminescence (PL) spectrum at each scan point. Automation ensures consistent step sizes, exposure settings, and timing. The Python scanning engine coordinates motion, checks stage stability, triggers spectral acquisition, and logs metadata for reproducibility. Preprocessing includes dark subtraction, baseline correction, and cosmic-ray removal using the double-acquisition method. Real-time visualization shows the evolving PL map and spectra, allowing for immediate adjustments.  

\subsection*{Fiber Alignment Scan}

To maximize fiber coupling, a raster scan of the Y-Z piezo voltages maps the coupling efficiency. The resulting heatmap displays a Gaussian-like peak indicating optimal alignment. 

\begin{figure}[h]
    \centering
    \includegraphics[width=0.5\linewidth]{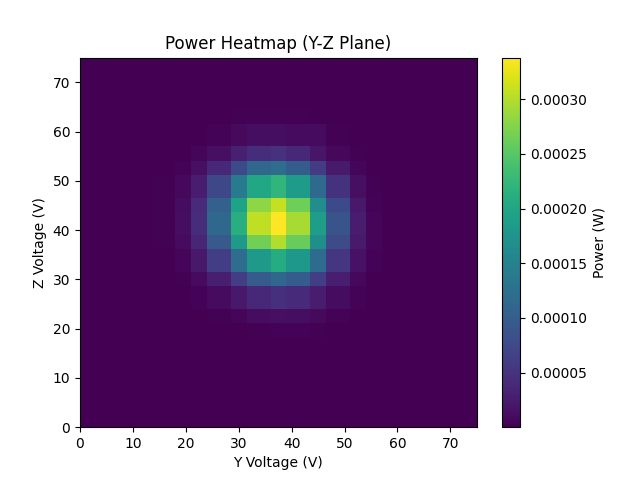}
    \caption{Fiber alignment scan showing coupling efficiency distribution.}
\end{figure}


\section{$g^{(2)}$ and Lifetime Measurement via snAPI}

Second-order autocorrelation determines if an emitter shows antibunching, a sign of single-photon emission. An Hanbury Brown and Twiss (HBT) setup splits PL onto two single-photon avalanche diodes (SPADs), and PicoQuant time-correlated single-photon counting (TCSPC) hardware records photon arrival timestamps. 

snAPI provides Python-level control of timing modes, detector thresholds, and data acquisition. In time-tagged time-resolved (TTTR) mode, timestamp streams from both detectors allow for detailed correlation analysis. To observe the sample's lifetime, we excite it with a pulsed laser and collect the emission in a SPAD, then plot the timestamps in a histogram. 

The laser intensity is set so that not too many photons are emitted. Receiving multiple photons during the pulsed laser period could cause a pile-up issue, resulting in undetected photons. For about 20 excitation pulses, we expect to detect one photon. We gather the timestamps to create our lifetime curve and fit it with a bi-exponential function to obtain our expected lifetime values. 

We have been unable to so far produce a material with a defect that has a PL emissive enough and isolated enough to produce an anti bunching signature in $g^{(2)}$ to display in this report.

The correlation function is computed as: 
\[
g^{(2)}(\tau) = \frac{\langle I(t) I(t+\tau) \rangle}{\langle I(t) \rangle \langle I(t+\tau) \rangle}
\]

Curve fitting of the anti-correlation dip reveals emitter purity.

\begin{figure}[h]
    \centering
    \includegraphics[width=0.5\linewidth]{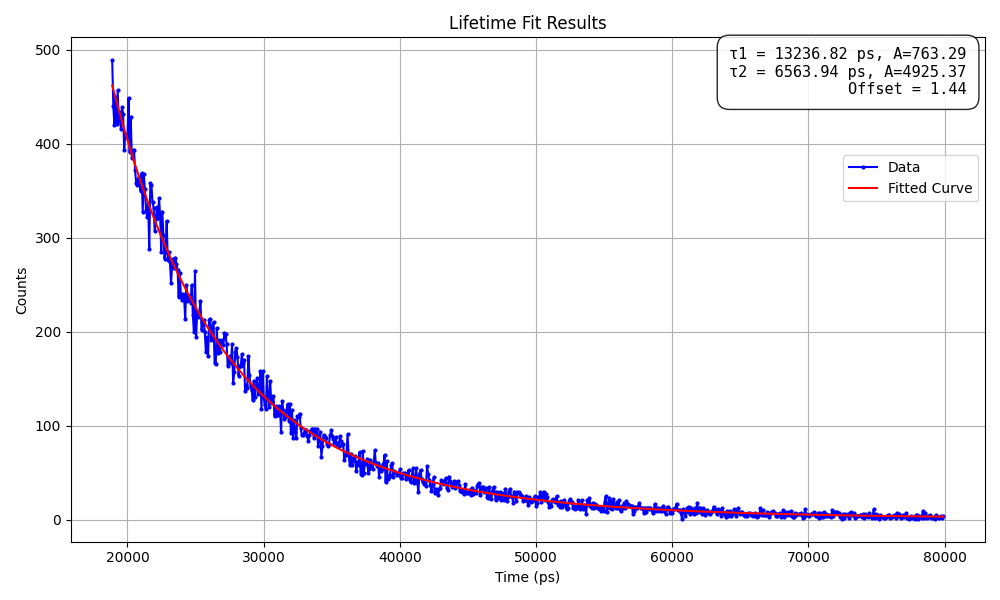}
    \caption{Lifetime decay curve of hBN Quantum Dot with fitted components.}
\end{figure}


\vspace{5cm}
\section{Galvo Mirror Implementation for Cryostat PL Mapping}

High-speed scanning is crucial for cryogenic PL mapping, where mechanical motion is restricted or vibration-prone. Galvo mirrors steer the beam into the objective at an angle and this induces a lateral translation movement in the plane of the sample. This characteristic is used to perform confocal mapping by changing the excitation spot without moving the sample stage.
\begin{figure}[h]
    \centering
    \includegraphics[width=1\linewidth]{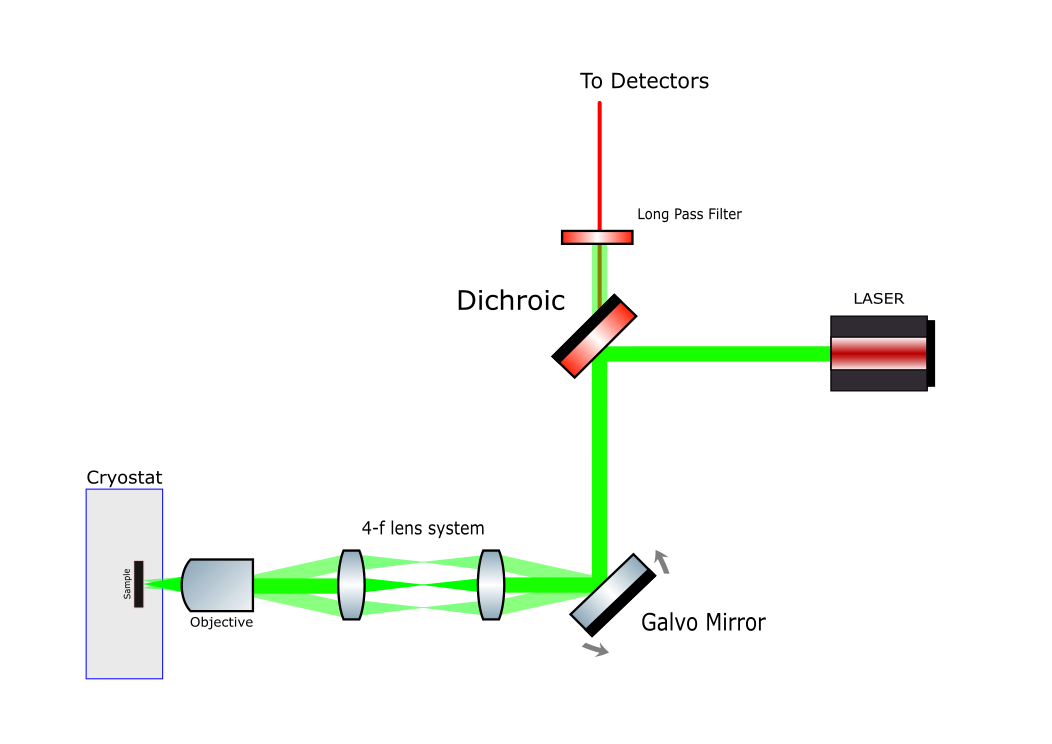}
    \caption{Galvo scanning setup schematic\cite{Hekmati2023}}
    \label{fig:placeholder}
\end{figure}
High-speed scanning is essential for cryogenic PL mapping, where mechanical motion is limited or prone to vibration. Galvo mirrors deflect the beam at an angle and a 4-f system is used to direct it into the objective at an angle, which causes a lateral movement in the sample plane. This feature allows for confocal mapping by shifting the excitation spot without moving the sample stage. The angle of the galvo mirrors is controlled through a DAQ input voltage. We then calibrate DAQ voltages to sample-plane coordinates: 
\[
x = \alpha_x V_x,\qquad y = \alpha_y V_y .
\]
Since galvos move faster than the spectrometer's integration time, software synchronizes the DAQ voltage updates, settling delays, and Andor exposure triggers. 

\begin{figure}[h]
    \centering
    \includegraphics[width=1\linewidth]{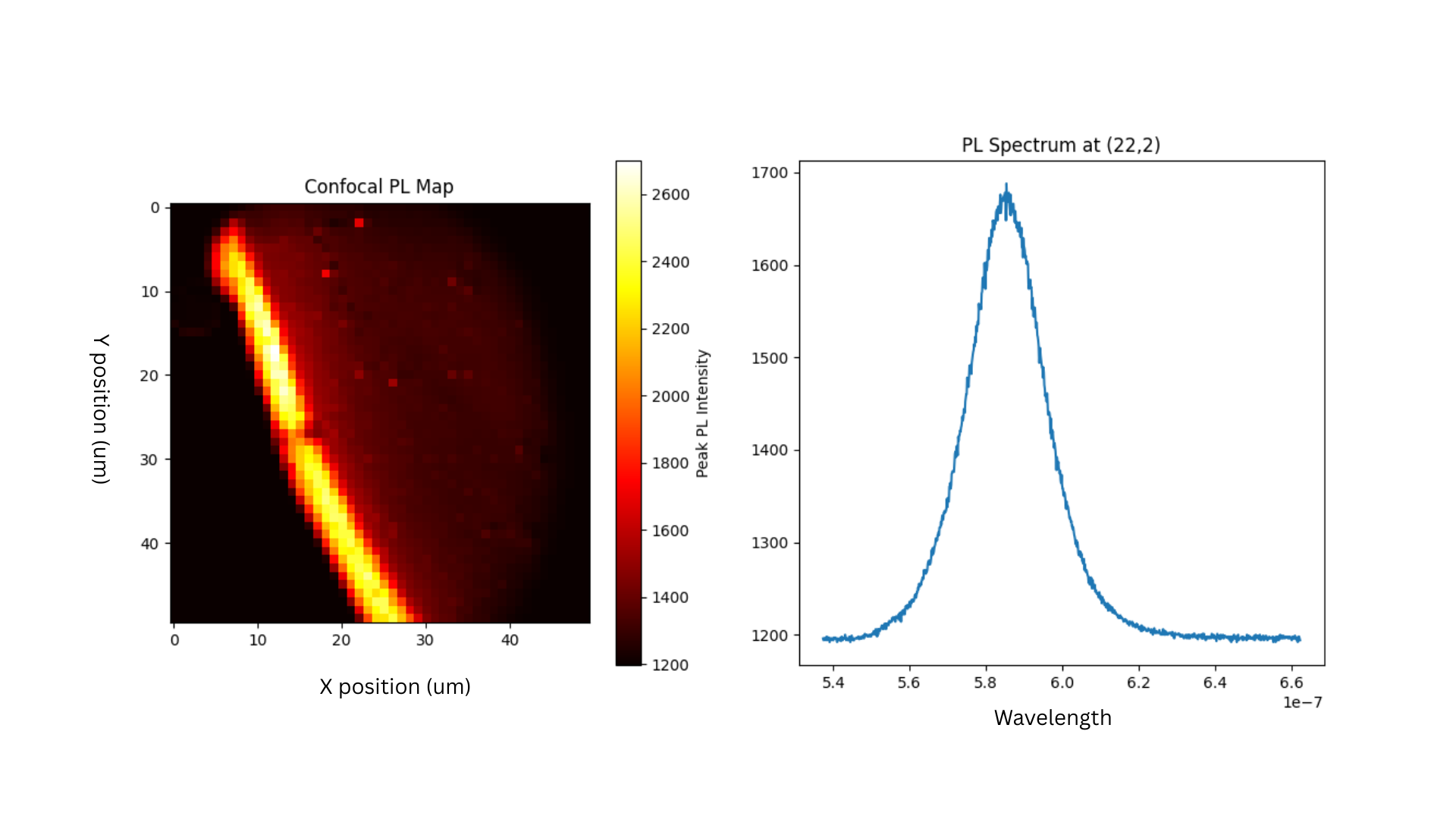}
    \caption{Confocal PL map of the edge of a spin coated CdSe Quantum Dot using galvo mirrors}
    \label{fig:placeholder}
\end{figure}

\section{Experiments with Common-Path Birefringent Interferometer}

\textbf{Nireos Gemini Interferometer:} We used a twin interferometer to generate optical path differences and convert them into wavelengths to get a spectrum of the transmitted light. We employed this device to capture hyperspectral images of our sample. Once this hyperspectral cube are obtained, we can isolate an image of a particular wavelength, where we expect our PL emission. This would ideally give us bright spots where our defect centers or emission centers exist. The process works as follows: 
\begin{itemize}
    \item Acquire images across the full range of stage or beam positions.
    \item Approximate the camera pixels to a $40 \times 40$ grid for testing to allow faster computation.
    \item Perform a discrete Fourier transform (DFT) on the position-dependent intensity of each pixel.
    \item Repeat this for all pixels to construct the full hyperspectral data cube.
\end{itemize}
\begin{figure}[h]
    \centering
    \begin{subfigure}[t]{0.49\linewidth}
        \centering
        \includegraphics[width=\linewidth]{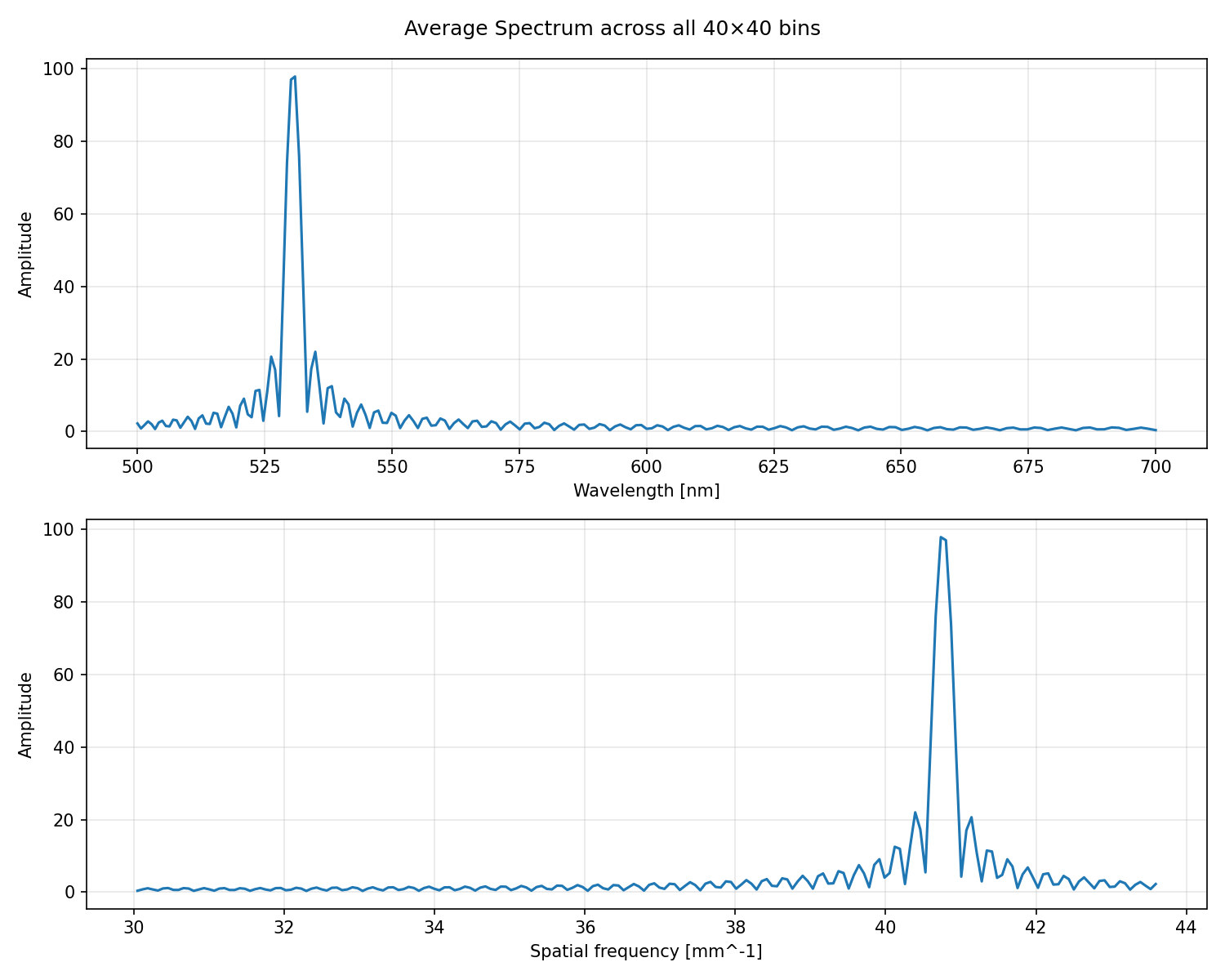}
        \caption{Average spectrum.}
    \end{subfigure}
    \hspace{0\linewidth} 
    \begin{subfigure}[t]{0.49\linewidth}
        \centering
        \includegraphics[width=\linewidth]{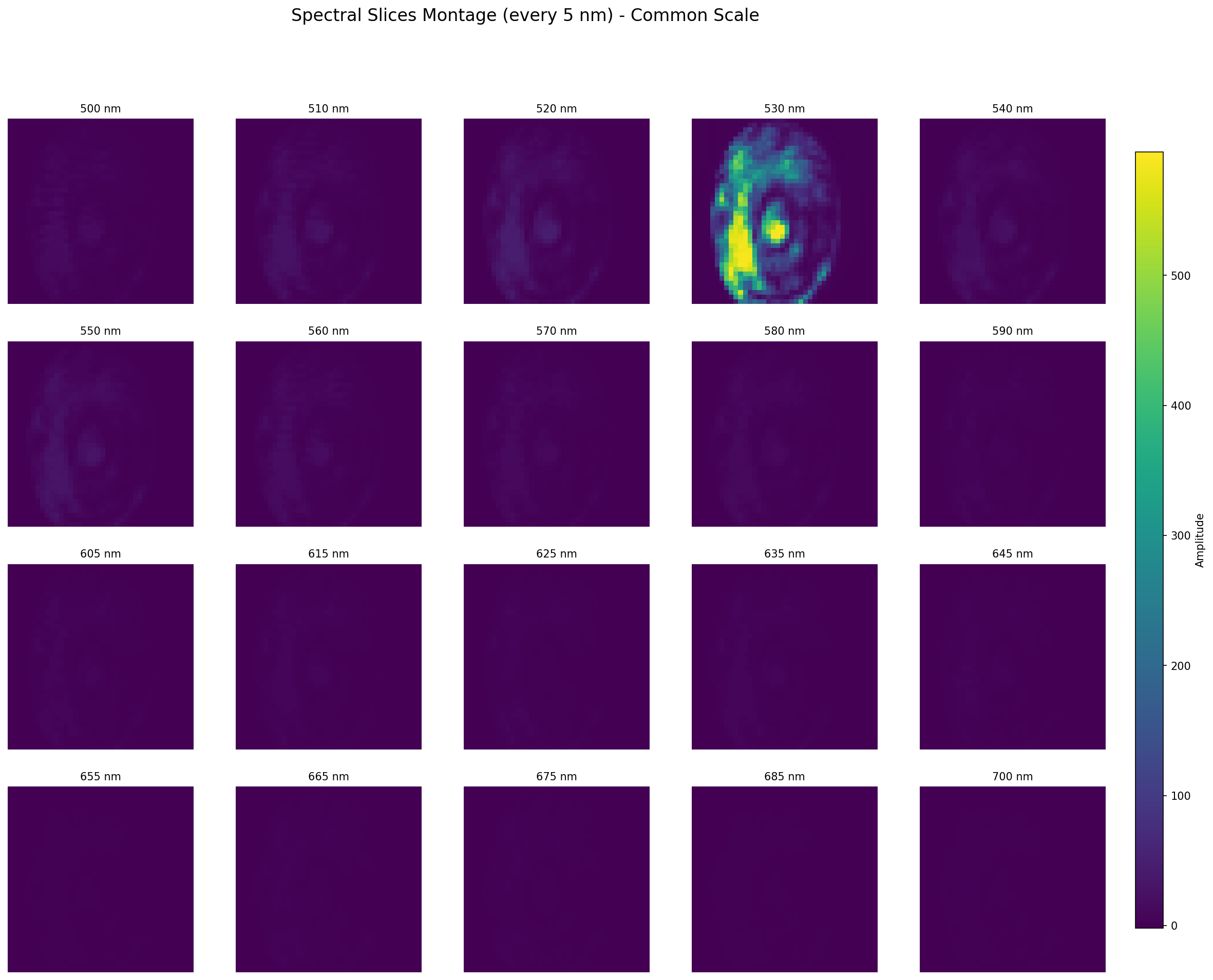}
        \caption{Spectral slices montage.}
    \end{subfigure}
    \caption{Hyperspectral pipeline outputs.}
\end{figure}


\section{Discussion and Challenges}

Several challenges emerged during automation:

\textbf{Cosmic-ray artifacts:} Random spikes in CCD spectra distort PL maps, which occur due to cosmic rays, which are high energy particles from space that occasionally hit the camera. The double-acquisition method effectively removes these artifacts by comparing two consecutive exposures at the same spatial coordinate and any stark differences in them that cross the provided threshold will be replaced by the lower intensity. \cite{Barton2019}

\textbf{Piezo-stage hysteresis:} Bidirectional serpentine scans introduced distortions due to direction-dependent stage deviation. Switching to single-direction raster scanning eliminated position errors.

\textbf{Drift and vibration:} Long scans in cryogenic systems suffer from drift and coupling instabilities. Using galvo scanning reduces scan times, so we can experience lower drift.


\section{Future Work}

\textbf{Unified GUI:} A cross-platform interface that integrates hardware control, acquisition monitoring, and analysis tools would streamline operation and reduce configuration errors. The GUI is envisioned to incorporate all applications into one unified Interface for simplifying operations.

\textbf{Database integration:} Centralized storage of spectra, PL maps, timestamps, and metadata would enhance reproducibility, enable cross-experiment comparison, and support emitter cataloging.

\textbf{Machine learning:} ML models could automatically identify emitters, filter background, and trigger adaptive high-resolution scans or $g^{(2)}$ measurements. This would significantly accelerate emitter discovery and reduce manual intervention.


\section{Conclusion}

We have developed a modular Python automation framework that combines control of spectrometers, stages, galvo mirrors, TCSPC hardware, and other devices for optical characterization.

The system automates PL mapping, stitched imaging, $g^{(2)}$ measurements, and high-speed scanning compatible with cryostats. This boosts reproducibility and throughput. The double-acquisition cosmic-ray removal ensures high-quality data.

We plan to expand the system with a unified GUI, database connectivity, and ML-assisted emitter localization. These updates will move the framework closer to fully automated, intelligent optical characterization.

\section{Acknowledgment}

We acknowledge funding support from the National Quantum Mission, an initiative of the Department of Science and Technology, Government of India and my special thanks to the members of Laboratory of Optics of Quantum Materials for their continued support.

\section {Code Availability}
The codes can be found in this repository as a downloadable python package: \href{https://github.com/loqm/phoqupy}{PhoQuPy}

\bibliographystyle{unsrt}
\bibliography{references}

@article{Esmann2024,
  author  = {Esmann, Martin and Wein, Stephen C. and Ant{\'o}n-Solanas, Carlos},
  title   = {Solid-State Single-Photon Sources: Recent Advances for Novel Quantum Materials},
  journal = {Advanced Functional Materials},
  volume  = {34},
  number  = {30},
  pages   = {2315936},
  year    = {2024},
  doi     = {10.1002/adfm.202315936},
  url     = {https://onlinelibrary.wiley.com/doi/abs/10.1002/adfm.202315936}
}

@article{Chalfoun2017,
  author  = {Chalfoun, Joe and Majurski, Michael and Blattner, Tim and Bhadriraju, Kiran and Keyrouz, Walid and Bajcsy, Peter and Brady, Mary},
  title   = {MIST: Accurate and Scalable Microscopy Image Stitching Tool with Stage Modeling and Error Minimization},
  journal = {Scientific Reports},
  volume  = {7},
  number  = {1},
  pages   = {4988},
  year    = {2017},
  doi     = {10.1038/s41598-017-04567-y}
}

@article{Barton2019,
  author  = {Barton, Sinead J. and Hennelly, Bryan M.},
  title   = {An Algorithm for the Removal of Cosmic Ray Artifacts in Spectral Data Sets},
  journal = {Applied Spectroscopy},
  volume  = {73},
  number  = {8},
  pages   = {893--901},
  year    = {2019},
  doi     = {10.1177/0003702819839098},
  url     = {https://doi.org/10.1177/0003702819839098}
}

@misc{KrammPythonAutomation,
  author = {Kramm, Arthur},
  title  = {Automation of Optical Setups Using Python},
  institution = {University of Groningen},
  year = {2023}
}

@phdthesis{Hekmati2023,
  author       = {Hekmati, Reza},
  title        = {Photon Emitters in Hexagonal Boron Nitride: Creating, Enhancing, and Controlling Optically Active Colour Centres},
  school       = {Cardiff University},
  year         = {2023},
  type         = {PhD Thesis},
  url          = {https://orca.cardiff.ac.uk/id/eprint/158238/1/2023HekmatiRPhD.pdf},
  note         = {Accessed through ORCA repository}
}

\end{document}